# Concerns in Software Development: A Systematic Mapping Study

Sandun Dasanayake, Jouni Markkula, Markku Oivo
Department of Information Processing Science
University of Oulu, Finland
{sandun.dasanayake, jouni.markkula, markku.oivo} @oulu.fi

## ABSTRACT

**Context:** Successfully addressing stakeholder concerns that are related to software system development and operation is crucial to achieving development goals. The importance of using a systematic approach to addressing these concerns throughout the software development life cycle is growing as more and more systems are employed to handle critical tasks. **Objective:** The goal of this study is to provide an overview of addressing concerns across the software development life cycle. **Method:** A systematic mapping study was conducted using a pre-defined protocol. Four digital databases were searched for research literature and primary studies were selected after a three round selection process conducted by multiple researchers. **Results:** The extracted data are processed and the results are reported from different viewpoints. The results are also analyzed against our research goals. **Conclusion:** We show that there is a considerable variation in the use of terminologies and addressing concerns in different phases of the software development life cycle.

## Categories and Subject Descriptors

D.2.9 [**Software Engineering**]: Management – *Life cycle*

## 1. INTRODUCTION

Every software system has a designated set of primary functions it was designed to perform. The work carried out by them is the reason to build the system in the first place. However, functionality alone will not guarantee the desired behavior of the system. Depending on the context, the system should maintain various characteristics apart from performing its basic functions. Even though there has been an asymmetrical emphasis on the functionality of the system, essentially, the system's utility is determined by both its functionality and its other characteristics [3]. Developing modern day software systems operated in critical domains is challenging because various concerns should be taken into consideration during the system development.

Concern is a general term that can be used to indicate any particular interest. However, in the context of software systems development, concern is used to denote an interest that pertains to the system's development, its operation, or any other aspect that is relevant to one or more stakeholders [4]. The broad scope and the intuitive nature of the definition allow any interest related to the system, including system functionalities, to be classified as a concern. However, since functionalities are system specific, only general system characteristics are considered as concerns during this study. Apart from the general concerns and the system functionalities, a system may still have various other concerns that are specific to the given system and its operational environment.

As we started studying software architecture design decision-making in the context of multiple concerns, we realized that the concepts related to concerns in the software engineering domain are broad and vague. They are open to significantly different interpretations from different viewpoints. The goal of this study is to overcome these challenges by mapping the state of the art in addressing concerns throughout the software development life cycle. Even though software development paradigms such as aspect-oriented software development have attempted to address concerns systematically, their influence in software development is still limited to certain phases. We have noticed that there are several terms that are used to refer to concerns in general, a type of concern, or an association with concerns. Therefor identifying terms used in different phases of the software development life cycle is crucial to achieving the above goal.

Every type of software development model, ranging from traditional life cycle models like the waterfall and the spiral models to evolutionary development models such as the agile software development model, deals with concerns during the life cycle. There are variations of software development life cycle phases, depending on the development model used. In this study, we use five main software development life cycle phases — requirement analysis, software design, implementation, testing, and maintenance — that are visible in most of the widely used software development models [9]. Performing a systematic mapping study has been recognized as a starting point that can help researchers in establishing a baseline for further research activities [5]. So, we aim to use the outcome of this study as a basis for our planned research in software architecture design decision-making.

This research paper presents a systematic mapping study carried out to identify various aspects of addressing concerns throughout the software development life cycle. After the introduction, section 2 describes the process of the mapping study from protocol development to classification and data extraction in detail. Section 3 showcases the results of the study and briefly analyzes them against the research questions, while section 4 discusses the possible threats to the validity of the study. Finally, section 5 concludes the paper highlighting the important elements of the study, the lessons learned and the possible extensions.

## 2. SYSTEMATIC MAPPING STUDY

A systematic mapping study can be considered as a lightweight systematic literature review with a different scope. Even though they can be used complementary to each other, there are differences in goal, process, breadth, and depth [5]. Providing a classification and conducting a thematic analysis are among the main goals of a systematic mapping study. While there is less emphasis on the quality of the literature, data extraction is also limited to data that is required for conducting a planned thematic analysis. A mapping study can also be conducted in a larger field, as there is no need of a detailed evaluation of the literature. Since our objective is to study research literature that covers the whole of software development life cycles and to classify and analyze them based on different viewpoints related to concerns, we think a systematic mapping study is well suited to our goals.

### 2.1 Mapping Study Process

The systematic mapping process described in [5] and systematic review guidelines provided in [6] are adapted to develop the process and the systematic mapping protocol followed in this study. The main steps followed during the study are:

1. Define research questions
2. Develop mapping protocol
3. Search for literature
4. Select primary studies
5. Create classification scheme
6. Extract data
7. Report results

The steps were sequentially followed during the study even though outcomes of some steps were refined iteratively using the outcomes of the later steps.

### 2.2 Research Questions

Research questions in mapping studies are usually less specific and also play a trivial role in shaping up a search strategy compared to systematic reviews. But still, research questions set the base for the study. Classification, analysis, and reporting are all built around the research question. The following research questions are defined in the context of this study.

- RQ1: What is the terminology used in association with concerns in software development?
- RQ2: What are the most considered concerns in software development?
- RQ3: How do the terminology and the concerns vary between different phases of the software development life cycle?

### 2.3 Mapping Protocol

By following a pre-defined protocol rigorously throughout the study, a systematic literature review increases the integrity of the study while reducing the possibility of bias. Since the followed protocol is also included in the study report, if necessary, it is possible to validate the study by replicating the steps in the protocol. Explicitly defining search processes and inclusion/exclusion criteria in the protocol and reporting it as a part of the outcome is one of the major characteristics that distinguish systematic reviews and mapping studies from traditional reviews [2]. Since a systematic mapping process doesn't necessary follow all the elements in systematic review protocols, we adapted the systematic review protocol provided by [6] to build a protocol for this study. The created protocol included goals of the research, research questions, search strategies, primary study selection including inclusion/exclusion criteria and selection processes, classification and data extraction mechanisms, and reporting. The protocol was reviewed several times using a pilot study based on a sample of forty research papers before conducting the actual study.

### 2.4 Search for Literature

Based on the recommendations [1] and the researchers' previous experience, four digital databases were selected as literature sources: ACM Digital Library, IEEE Xplore, ScienceDirect, and Scopus. The next step of the search process was formulating the search strings. We made several attempts to build search strings using different terms related to the research area. After piloting many terms, we realized some of them gave an insufficient amount of results while others caused a large amount of results that could not be handled within the scope of this study. Since identifying terminologies and all the concerns involved was one main goal of the study, restricting our search to only a limited number of known terms would hinder the process. So we decided to conduct the search as broadly as possible and filter irrelevant papers during the selection process. Finally, we agreed to use both "software development" and "concern" as search terms because they gave a manageable amount of results in a broader scope. No limitations were applied about the published forums, as doing so would have limited the scope of the literature. But since our study was based in the software engineering domain, when the searching interface supported it, the search was restricted only to the software engineering domain. No restrictions were applied regarding the publication year.

### 2.5 Primary Study Selection

After conducting the search, the results were uploaded to Refworks [8]. The primary study selection was carried out in three rounds. In each round, two researchers used a set of selected sections of the papers available in the round and marked each paper as either "accept," "reject," or "can't decide" based on the pre-defined inclusion/exclusion criteria shown in Table 1. The inclusion/exclusion criteria were designed to be narrowed down in each round as more information became available.

**Table 1. Inclusion/exclusion criteria used during selection**

| Selection Process | | | Inclusion criteria *(If the article)* | Exclusion criteria *(If the article)* |
|---|---|---|---|---|
| 1st Round (metadata) | 2nd Round (titles) | 3rd Round (abstracts) | is a conference paper | is not written in English |
| | | | OR is a journal paper | OR is a duplicate |
| | | | OR is a book chapter | OR is a presentation |
| | | | OR is a technical report | OR is a interview or a panel discussion |
| | | | OR is a Ph.D. thesis | OR is a summary / extended abstract |
| | | | OR has a scope that lies within the software engineering domain | OR is a master's thesis. |
| | | | AND is related to software development | - |
| | | | AND uses the term "concern" in the context of the definition used in this study | OR is about the same study on which another more recent paper accepted in this round is based |

The first round of selection was done using the metadata of each paper collected during the search. The duplicates, the content types that weren't considered in this study, and the papers that were from the sources that were not relevant to the software engineering domain were filtered out in this round. RefWorks was used to identify duplicates, but the identified papers were manually checked before removal to avoid false flags. The researchers went through the titles of the papers in the next round and the abstracts in the third rounds.

Once all the papers in the round were marked independently by each researcher, the lists were combined and the decision to include or exclude the paper was made based on the combined result. When both researchers agreed to accept or reject a paper, or one of the researchers marked as "can't decide" while the other made a decision, the acceptance or rejection of the paper was made right away. If they disagreed or both of them couldn't decide, it was marked as a conflict. After each round, a conflict resolution meeting was held to resolve conflicts. If the researchers were unable to agree on a paper, it was carried into the next selection round along with the selected papers. After the third and the final round, a third researcher participated in the conflict resolution meeting to resolve the remaining conflicts.

## 2.6 Classification and Data Extraction

Developing proper classification schema helps not only to extract correct data but also to report the findings of the study in an effective manner. Based on the research questions addressed in this study, three classification schemes were derived: phase of the software development life cycle, types of concerns addressed in the research, and terms used to refer to concerns. Apart from the main classification schemes, other classification schemes also emerged while conducting the study.

Based on the classification scheme, the following data were extracted from selected research papers for analyzing and reporting: title of the paper, author(s), year of publication, type(s) of concerns addressed, phase(s) of software development life cycle, and term(s) used to refer to concerns. The extracted data were stored in an Excel file for further processing. To make sure that the data extraction was done properly, two researchers independently performed data extraction on a selected sample of 100 papers, which was 24% of the total amount of papers. The results were compared to identify if there was any inconsistency. Since there were no major differences between the results, a single researcher completed data extraction from the rest of the research papers based on the agreed process.

## 3. RESULTS

The main goal of reporting results is to answer the research questions and to uncover other possible information. At the same time, by combining the described protocols with the data that are presented, it is possible to validate that the study has been conducted as claimed.

## 3.1 Selection Process Results

The initial search was conducted on the four selected databases using the search strings that are defined based on the scope of the study. As shown in Table 2, it brought 2711 papers that contain the given search strings in the title, the abstract, or the keywords. Scopus had the largest number of papers while ScienceDirect had the smallest. At the end of the three selection rounds, 421 papers, 16% of the initial search results, were selected as the primary studies for data extraction. The list of the selected primary studies is available here: http://dasanayake.com/EASE2014.pdf

**Table 2. Paper count after each selection round**

| Database | Initial | 1st | 2nd | 3rd |
|---|---|---|---|---|
| ACM Digital Library | 331 | 266 | 255 | 101 |
| IEEE Xplore | 941 | 797 | 616 | 189 |
| ScienceDirect | 116 | 92 | 57 | 20 |
| Scopus | 1323 | 594 | 379 | 111 |
| Total | 2711 | 1612 | 1307 | 421 |
| Percentage | 100% | 59% | 48% | 16% |

## 3.2 Extracted Results

Information extracted during the study is presented and analyzed from different viewpoints to provide the answers to the research questions.

### 3.2.1 Terminology used

As mentioned earlier, identifying terminology associated with concerns is one of the goals of the study. During the study, it was noted that while some research papers used multiple terms to refer to concerns, some papers only used the term "concern." Another observation was that a single term was used to express different meanings in different papers. For example, even though the majority of papers used the term "aspect" to refer to cross-cutting concerns, there were many papers where it was used to refer to concerns in general or some other type of concern.

According to the results from the study, "aspect" was the most common term used to associate concerns. The second most frequently used term was "non-functional requirement." Since it is a commonly used term in software engineering, it was expected to have a bigger study count than the count resulted. Its lack of association with the term "concern" can be seen as a reason for that. "Quality attribute" was the other term that occurred frequently. While there were many terms that were used less frequently, most of them were different combinations of words such as quality, non-functional, aspect, property, or attribute.

### 3.2.2 Concerns addressed

As expected, we came across a variety of concerns during the study. While most of them are common qualities of software systems, there were many software system specific concerns also. Since some studies had used the term "concern" to refer to functional properties of the system, most of the software system specific concerns were functional concerns. As shown in Table 3, security, reusability, and maintainability are the most commonly considered concerns in the selected studies. Evolvability and performance are also among the top 5 concerns. Concerns such as reusability, maintainability, traceability, and modularity can be placed in a special category, as they are the concerns that can be addressed by means of separation of concerns.

**Table 3. Top 5 Concerns addressed**

| Concern | Study count |
|---|---|
| Security | 49 |
| Reusability | 48 |
| Maintainability | 25 |
| Evolvability | 17 |
| Performance | 14 |

*3.2.3 Variation in the software development life cycle*

Figure 1 shows a compact overview of addressing concerns throughout the software development life cycle. As shown there, design and implementation are the phases where most of the concerns are addressed. Active use of aspect-oriented software development paradigm during the design and the implementation can be the main reason to address most of the concerns in these stages. Since the design is an early phase of the software development life cycle, if concerns are handled in the design phase, it affects most parts of the life cycle. The variation of terminologies and specific concerns between different phases are largely proportional to the overall concern variation, and they will be further analyzed in a detailed study.

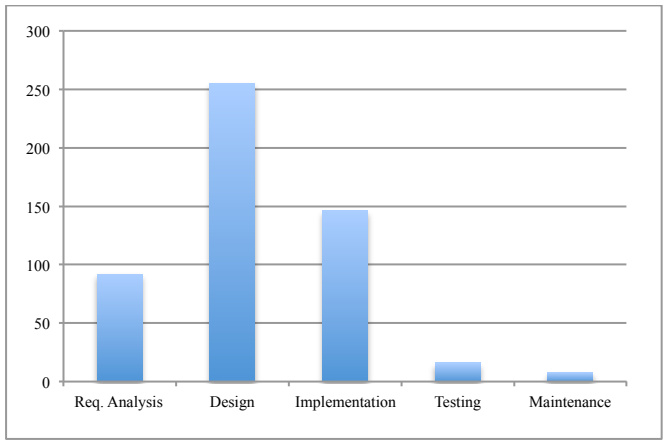

**Figure 1. Concerns in software development life cycle phases**

## 4. THREATS TO VALIDITY

Despite following a rigorous process throughout the study, there are some issues that can be considered as threats to the validity of the research. Our search term selection can be seen as a drawback of the study. Rather than limiting to the term "concern," we could have used terms, such as "aspects," "non-functional requirements," and "quality attributes," that described concerns or some aspects of concerns. Even though doing so may enhance the search result quality, at the same time, it could also bring an unprecedented number of search results that cannot be handled within the scope of this study. So, we had to make a trade-off between using those terms and covering the whole of software development life cycles. Having a large number of primary studies may also pose a threat to the validity of the research as it makes the extraction process prone to errors. To minimize that threat, two researchers independently performed data extraction on a selected sample and the identified issues were clarified before performing the final data extraction.

## 5. CONCLUSION

This study was conducted as the first step of a research focused on software architecture design decision-making in the context of multiple concerns. The goal of this study was to provide an overview of addressing concerns in different phases of the software development life cycle while specially focusing on the terminology and type of concerns. So, the results of the study could be used as a basis for building a unified view of the concerns. The study answered the research questions by providing an overview about the terminology used in association with concerns, the most considered concerns and their variation in the software development life cycle. Apart from achieving the desired goals, many other results were also produced during the study. Specially, several lessons were learned related to the research area as well as the process of conducting systematic literature studies.

While conducting the study, we focused on two main areas that were recognized as the most important elements of a successful systematic mapping study. One area was the results, and the other was the process. Though the primary goal of the study was to answer the research questions, the validity of the research would be in doubt if a proper systematic process were not followed during the study. On the other hand, following the best possible process would not be useful without obtaining interesting results. We tried to find the balance between those two elements by following a rigorous systematic process while keeping our focus on the study goals. Finally, the results produced in this study can be utilized to conduct new studies focused on different concerns and phases of the software development life cycle.

## 6. ACKNOWLEDGMENTS

This work is funded by ITEA2 and Tekes - The Finnish Funding Agency for Technology and Innovation, via the MERgE project. We would also like to thank Prof. June Verner for the guidance and the feedback provided during this study.